\documentclass[aps,prl,twocolumn, showpacs]{revtex4}

\usepackage{graphicx}

\begin{document}
\draft

\title{Restrictions on modeling spin injection by resistor networks}
\author{Emmanuel I. Rashba}
\address{Department of Physics and Center for Nanoscale Systems, Harvard University, Cambridge, Massachusetts 02138, USA\\
and Department of Physics, Loughborough University, Leicestershire LE11 3TU, UK}

\begin{abstract}
Because of the technical difficulties of solving spin transport equations in inhomogeneous systems, different resistor networks are widely applied for modeling spin transport. By comparing an analytical solution for spin injection across a ferromagnet - paramagnet junction with a resistor model approach, its essential limitations stemming from inhomogeneous spin populations are clarified.
\end{abstract}
\maketitle


Conventional electronics is based on a single parameter of electron, its charge. Therefore, electronics deals only with electron trajectories. It does not involve electron spin, its internal degree of freedom. The new paradigm of spin-based electronics, or spintronics, is based on active involvement of electron spin in transport and optical phenomena, and on employing electron spin for both information processing and information storage. During the last decade, semiconductor spintronics developed into a wide and diversified field. Early review papers \cite{Prinz,Wolf} were followed by more recent surveys covering specific scientific problems and technological perspectives of this rapidly developing field \cite{Awsch02,Gregg,Ohno,GanPr,Silsbee,Zutic,Schmidt,Tserkov,Engel,Flatte,Jake,HansonRMP,Sinitsyn,Dietl,Bratk}. There exists close connection between the recent work on electric spin manipulation in low-dimensional systems and the previous work on anomalous Hall effect \cite{KarLut}, electric dipole spin resonance \cite{EDSR}, optical orientation \cite{OptOr} and photogalvanic effect \cite{IvPik,BelinStu} in three-dimensional systems. Strong impetus for semiconductor spintronics was given by the discovery of giant magnetoresistance in metallic systems \cite{Baibich,Binash} and by its impressing practical applications.

Spin injection from ferromagnetic sources into paramagnetic media is believed to be an important part of the new phenomena and applications in the field of the spin-polarized electron transport. The concept of a field effect spin transistor \cite{Datta} served as one of the stimuli. Successful experiments on spin injection into superconductors \cite{TedMes} and normal metals \cite{JohnSil} were very promising, and theoretical work supported reliability of the idea \cite{AronPik,Aronov,JohnSil87,Son,Valet,Hershfield}. Successful detection of the floating potential (spin-e.m.f., electromotive force \cite{Silsbee80}) at a ferromagnetic probe became an independent confirmation of efficient spin injection. Meantime, the methods that resulted in a remarkable success in spin injection from ferromagnetic metals into paramagnetic metals turned inefficient as applied to the spin injection from ferromagnetic metals into semiconductors. Spin injection at the level of only about 1\% was reported \cite{Hammar99,Gardelis}, and persuasiveness of the results was disputed \cite{Monson,Wees}. Absence of any measurable spin-e.m.f. also indicated that the concentration of the electrically injected nonequilibrium spins was vanishingly small \cite{Filip}.

Inefficiency of a ``perfect" contact between a ferromagnetic metal and a semiconductor as a spin emitter, that seemed puzzling, found its natural explanation in the framework of the conductivity mismatch concept \cite{Schmidt00}. The next step was proposing resistive spin selective contacts, like tunnel or Schottky barriers, as spin sources \cite{Rashba00}. The underlying physics is as follows. Elements of the circuit with small effective resistances are ``soft", i.e., the concentration of nonequilibrium spins in them adjusts to the regime imposed by the elements with large effective resistances. E.g., a ferromagnetic metal connected to a semiconductor by a ``perfect" (zero-resistance) contact is such a soft element. Inside it the diffusive current of minority spins enhances their Ohmic current, while the diffusive current of majority spins partly compensates their Ohmic current. As a result, electric current across the contact becomes nearly spin unpolarized, hence, good metal is a poor spin emitter. However, both the metal and semiconductor become ``soft" elements if the resistance of a barrier inserted between them is larger than their effective resistances. Under these conditions, spin injection is controlled by spin selectivity of the barrier, i.e., by the difference in its resistances for up- and down-spin electrons.

More detailed diffusive theories of spin injection confirmed this concept \cite{FertJaf,Rashba02}, and it was generalized for ballistic transport across spin valves \cite{Kravch,Jiang,Rashba03} and for optimization of Schottky barriers \cite{Albrecht}. It also explained dramatic inefficiency of low-resistance contacts, and early observations of spin injection from STM tips \cite{Alvorado}, resonant double barriers \cite{Ohno98}, and Schottky barriers \cite{Monsma}. Finally, it allowed increasing spin injection coefficients from ferromagnetic metals into semiconductors to the level of dozens of percents nearly immediately \cite{Zhu,Hammar01,Hanbicki,Motsnyi,Kreuser,Hu}. Resistive contacts became instrumental even in increasing the metal-to-metal spin injection \cite{Jedema}, i.e., in a system where no considerable conductivity mismatch could be anticipated.
Actually, the conductivity mismatch (or ``bottlenecking") can already been found in some previous theories \cite{JohnSil87,Hershfield}, but importance of it has not been properly recognized then.

Different approaches for overcoming the suppression of spin injection by conductivity mismatch are based on developing semimagnetic semiconductors \cite{Oester99,Osipov,Fiedr99,Ohno99,Jonker00,Dietl00} and half-metals \cite{Pick01,Park02,Mav04} for spin emitters.

The concepts discussed above are based on the solutions found for planar, i.e., one-dimensional, geometry. It is a price for deriving the results that are exact. The limitations of this geometry are very restricting. Geometry of real experimental systems is much more complicated, and it is well known that specific geometry around contacts influences spin injection \cite{Raichev}. Any deviation from planar geometry requires applying either approximate \cite{TakaMa} or numerical methods \cite{JohnBY}. Even for a strictly planar geometry only a F-N-junction, i.e., a junction between a ferromagnet and a normal conductor (paramagnetic metal or semiconductor) can be easily solved. Solving a diffusive F-N-F valve is elementary but cumbersome. Using a proper calculational technique allows simplifying the procedure essentially \cite{Rashba02}, but a discrepancy in the results still persisting reflects existence of technical problems.

Under such conditions, in designing devices and analyzing their behavior in different regimes various resistor network schemes are widely used. An excellent example of exploiting such an approach can be found in the review paper by Schmidt \cite{Schmidt}. Nevertheless, a general question exists: {\it in which extent applying such network schemes to the geometries that can be solved exactly proves the consistency of the network approach?} Also, which level of accuracy can be expected and which factors influence it? It is shown in what follows that for a F-N-junction in a linear (Ohmic) regime the spin injection coefficient and junction resistance cannot be described in a framework of a single resistor network scheme.

Spin injection coefficient $\gamma$ is defined as
\begin{equation}
\gamma=(I_\uparrow-I_\downarrow)/I\,,\,\,I=I_\uparrow+I_\downarrow\,,
\label{eq1}
\end{equation}
where $I_\uparrow$ and $I_\downarrow$ are the currents of spin-up and spin-down electrons, respectively, at the center of the junction (F-N interface), and $I$ is the total electric current. For a resistive F-N-contact, the expression for $\gamma$ is well known \cite{Hershfield,Rashba00}
\begin{equation}
\gamma=[r_c(\Delta\Sigma/\Sigma)+r_F(\Delta\sigma/\sigma_F)]/r_{FN}~,
\label{eq2}
\end{equation}
where
\begin{equation}
r_{FN}=r_F+r_c+r_N,~~r_c=\Sigma/4\Sigma_\uparrow\Sigma_\downarrow\,.
\label{eq3}
\end{equation}
Here $r_F$ and $r_N$ are effective (diffusive) resistances of the ferromagnet and the normal conductor
\begin{equation}
r_F=\sigma_FL_F/4\sigma_\uparrow\sigma_\downarrow,~~r_N=L_N/\sigma_N\,,
\label{eq4}
\end{equation}
$L_F$ and $L_N$ are spin diffusion lengths in them, and
$\sigma_\uparrow$ and $\sigma_\downarrow$ are the
conductivities of spin-up and spin-down electrons,
respectively, in the ferromagnet. Then
$\sigma_F=\sigma_\uparrow+\sigma_\downarrow$ is the total
conductivity of the ferromagnet,
$\Delta\sigma=\sigma_\uparrow-\sigma_\downarrow$ describes spin
polarization in the bulk of the ferromagnet, and $\sigma_N$ is
the conductivity of the N-conductor. As distinct from these
bulk parameters, $\Sigma_\uparrow$ and $\Sigma_\downarrow$ are
the conductivities of the tunnel barrier for spin-up and
spin-down electrons, respectively, with
$\Sigma=\Sigma_\uparrow+\Sigma_\downarrow$ and
$\Delta\Sigma=\Sigma_\uparrow-\Sigma_\downarrow$. Evidently,
the ratio $\Delta\Sigma/\Sigma$ measures spin selectivity of
the resistive contact. When deriving Eq.~(\ref{eq1}) it was
assumed that the contact is spin conserving, i.e., $I_\uparrow$
and $I_\downarrow$ are continuous functions of $x$ at $x=0$,
FIG.~1(a). It is seen from Eq.~(\ref{eq3}) that $r_c$ has a
meaning of an effective resistance of the contact, that
together with $r_F$ and $r_N$ makes the total effective
resistance of the junction $r_{FN}$. For $r_c=0$, equation for
$\gamma$ reduces to
$\gamma=r_F(\Delta\sigma/\sigma_F)/(r_F+r_N)$
\cite{Son,Schmidt00}.

Schmidt {\it et al.} \cite{Schmidt01} and Jonker {\it et al.} \cite{Jonker03} proposed simple and appealing resistor networks for systems including magnetic and normal conductors, see FIG.~1(b). Despite the fact that both networks are ``topologically equivalent" and look as nearly identical, they were proposed for different systems. Therefore, the meaning of the resistances in these networks is rather different. The network of Ref.~\onlinecite{Schmidt01} was proposed for a F-N-F spin valve with infinite spin flip time in the normal region, $L_N^{-1}=0$, and is unrelated to the basic subject of the present paper. On the contrary, the network of Ref.~\onlinecite{Jonker03} was proposed for spin injection across a F-N-junction and is an excellent candidate for comparing with the results derived for diffusive regime.

The network of Ref.~\onlinecite{Jonker03}, see FIG.~1(b), consists of two channels, each of them for a single component of electron spin, $\alpha=\uparrow,\downarrow$. Inside each channel, effective resistances of F- and N-conductors and of the barrier contribute in series as
\begin{equation}
R_\alpha=L_F/\sigma_\alpha+1/\Sigma_\alpha+2L_N/\sigma_N\,.
\label{eq5}
\end{equation}
Because the ratio of the currents across these channels equals $I_\uparrow/I_\downarrow=R_\downarrow/R_\uparrow$, one recovers Eq.~(\ref{eq2}) after simple algebra. Therefore, the resistor network of FIG.~1(b) describes spin injection coefficient $\gamma$ perfectly, as it was already stated by Jonker {\it et al.} \cite{Jonker03}. Next physical quantity of interest is the electrical resistance $R$ of a F-N-junction. For the resistor network of FIG.~1(b) it equals
\begin{equation}
R_\gamma^{-1}=R^{-1}_\uparrow+R^{-1}_\downarrow\,.
\label{eq6}
\end{equation}

Electrical resistance $R$ of a diffusive F-N-junction was found in Ref.~\onlinecite{Rashba00}, and detailed derivation was provided in Ref.~\onlinecite{Rashba02}. However, because this resistance is critical for our final results, we outline here the basic guidelines for the derivation. On both sides of the junction, inside the F- and N-conductors, electrochemical potentials $\zeta_{\uparrow,\downarrow}(x)$ of up- and down-spin electrons obey the standard drift-diffusion equations \cite{JohnSil87,Son,Valet,Hershfield}. The contact conductivities $\Sigma_{\uparrow,\downarrow}$ of the F-N-interface, $x=0$, are defined by the boundary condition
\[j_{\uparrow,\downarrow}(0)=\Sigma_{\uparrow,\downarrow}(\zeta_{\uparrow,\downarrow}^N(0)-
\zeta_{\uparrow,\downarrow}^F(0))\,,\]
where $j_{\uparrow,\downarrow}(x)$ are current densities of up- and down-spin electrons, respectively. The superscripts N and F in $\zeta^{F,N}(0)$ indicate that the corresponding potentials $\zeta_{\uparrow,\downarrow}(0)$ should be taken at the F- and N-sides of the interface, respectively. This boundary condition implies the continuity of $j_{\uparrow,\downarrow}(x)$ at $x=0$, what is tantamount to spin conservation at the F-N-interface. We note that Eq.~(\ref{eq2}) was derived under the same conditions and by the same procedure.

Under these conditions, the F-N-junction resistance $R$, excluding nominal resistances of the F- and N-conductors, equals  \cite{Rashba00,Rashba02}
\begin{eqnarray}
 R=\Sigma^{-1}&+&{1\over{r_{FN}}}\{ r_N\left[r_c(\Delta\Sigma/\Sigma)^2
+r_F(\Delta\sigma/\sigma_F)^2 \right]\nonumber\\
&+&r_c r_F\left[(\Delta\Sigma/\Sigma)-(\Delta\sigma/\sigma_F)\right]^2 \}\,.
\label{eq7}
\end{eqnarray}
Remarkably, the resistance $\Sigma^{-1}$ of a tunnel barrier appearing here differs from the effective resistance $r_c$ of Eq.~(\ref{eq3}). Second term in (\ref{eq7}) represents the nonequilibrium part of $R$ that is always positive. It vanishes when $L_F,L_N\rightarrow0$ and nonequilibrium spin populations cease to exist; under these conditions $r_F,r_N\rightarrow0$. F-N-junction acquires a finite resistance, $R\neq0$, even for a zero-resistance barrier, $r_c=\Sigma^{-1}=0$. We note that throughout this paper the term ``resistive contact" implies $r_c\,,\Sigma^{-1}\neq0$.

Comparing Eqs.~(\ref{eq6}) and (\ref{eq7}) shows that $R\neq R_\gamma$. They are related by the equation
\begin{equation}
R+L_F/\sigma_F+r_N=R_\gamma\,.
\label{eq8}
\end{equation}
It shows that $R_\gamma> R$, and $R$ and $R_\gamma$ differ by the diffusive resistances of the ferromagnet, $L_F/\sigma_F$, and the normal conductor, $r_N$ [of which the former does not coincide with $r_F$ of Eq.~(\ref{eq4})]. Eq.~(\ref{eq8}) is shown graphically in FIG.~1(c).

It is seen from Eq.~(\ref{eq8}) that the difference between $R$ and $R_\gamma$ is a general phenomenon. It is related to spin diffusion in F- and N-regions rather than to a tunnel barrier separating them. Therefore, to uncover the underlying physics we consider a transparent barrier ($r_c\,,\Sigma^{-1}=0$) when Eqs.~(\ref{eq6}) and (\ref{eq7}) are essentially simplified. It follows from Eq.~(\ref{eq7}) that in this case electrical resistance of the junction equals
\begin{equation}
R={{r_Fr_N}\over{r_F+r_N}}\bigg({{\Delta\sigma}\over{\sigma_F}}\bigg)^2\,.
\label{eq*}
\end{equation}
Therefore it vanishes, $R=0$, whenever one of the spin diffusion lengths, either $L_F$ or $L_N$, vanishes. It is a result that is anticipated from simple physical arguments. Indeed, when $r_c\,,\Sigma^{-1}=0$, electrochemical potentials of up- and down-spin electrons, $\zeta_{\uparrow,\downarrow}(x)$, are continuous at $x=0$
\begin{equation}
\zeta^F_\uparrow(0)=\zeta^N_\uparrow(0)\,,\,\,\zeta^F_\uparrow(0)=\zeta^N_\uparrow(0)\,.
\label{eq9}
\end{equation}
When one of the diffusion lengths, say $L_N$, vanishes, electron spins are in equilibrium in the whole N-region, hence, $\zeta^N_\uparrow(x)=\zeta^N_\downarrow(x)$ for $x\geq0$. Then it immediately follows from Eq.~(\ref{eq9}) that $\zeta^F_\uparrow(0)=\zeta^F_\downarrow(0)$, and this suggests that $\zeta^F_\uparrow(x)=\zeta^F_\downarrow(x)$ in the whole F-region, $x\leq0$. Therefore, spin equilibrium on one side of the junction maintains spin equilibrium also on the other side of it, and second term in Eq.~(\ref{eq7}) always vanishes in the absence of spin nonequilibrium. Under these conditions, $R=\Sigma^{-1}$.

One can check by inspection that $R_\gamma$ of Eq.~(\ref{eq6}) does not obey this requirement, and it vanishes only when both spin diffusion lengths vanish, $L_F=L_N=0$. Presence of second and third terms in the left hand side of Eq.~(\ref{eq8}) reflects this critical difference in the properties of $R$ and $R_\gamma$. Therefore, the network of FIG.~1(b) provides perfect description of spin injection coefficient $\gamma$, but cannot provide a consistent description of the junction resistance $R$. When barrier resistance $\Sigma^{-1}$ is high, $R_\gamma$ describes consistently the leading term in $R$. Inaccuracy in the next terms, originating from spin imbalance in the bulk, depends on specific parameter values.

In conclusion, it is instructive to compare two-channel resistor models of spin injection with the Mott's two-channel model of spin transport in ferromagnetic metals \cite{Mott36}. In the latter model, scattering influences only the mobilities of the carriers with different spins. Because of homogeneity of the system, scattering does not produce spin imbalance and nonequilibrium spin populations. That is why the conductivities $\sigma_\uparrow$ and $\sigma_\downarrow$ are well defined. On the contrary, nonequilibrium spin populations are central for the spin injection problem, and all processes responsible for spin relaxation critically influence spacial distribution of these populations. Two-channel resistor models, at least in their simplest realizations, are not properly fit for describing the effect of inhomogeneous populations.


\vspace{1cm}



\vspace{3mm} \providecommand{\figdim}{3.375in}
\providecommand{\suff}{mod.pdf}

\begin{figure}
\begin{tabular}{lr}
\includegraphics[width=1.8in]{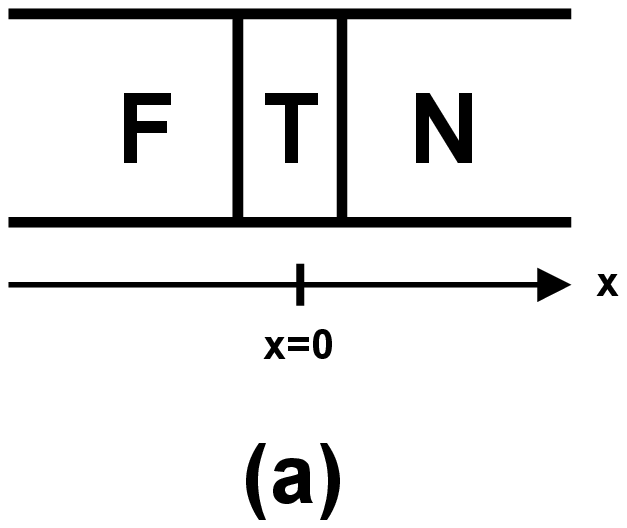}\\
\includegraphics[width=\figdim]{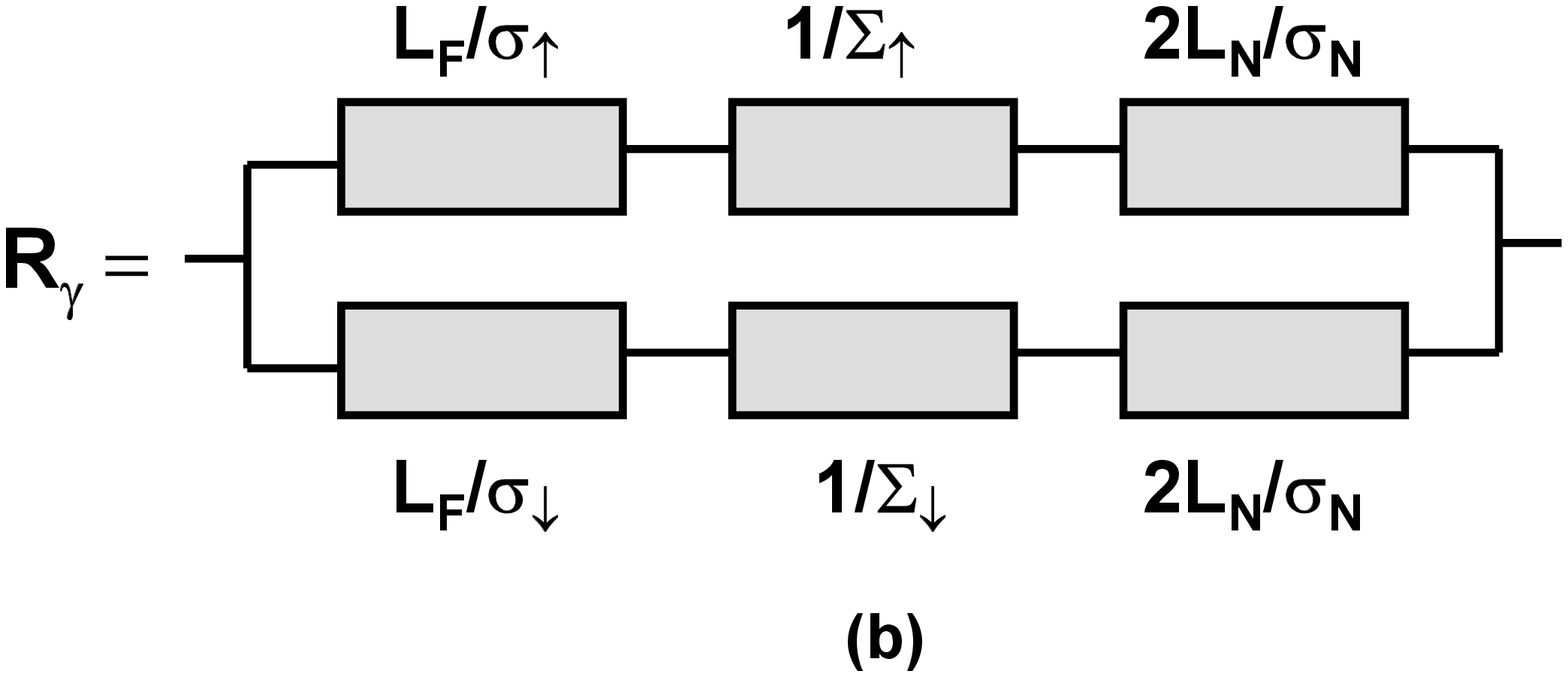}&
\includegraphics[width=\figdim]{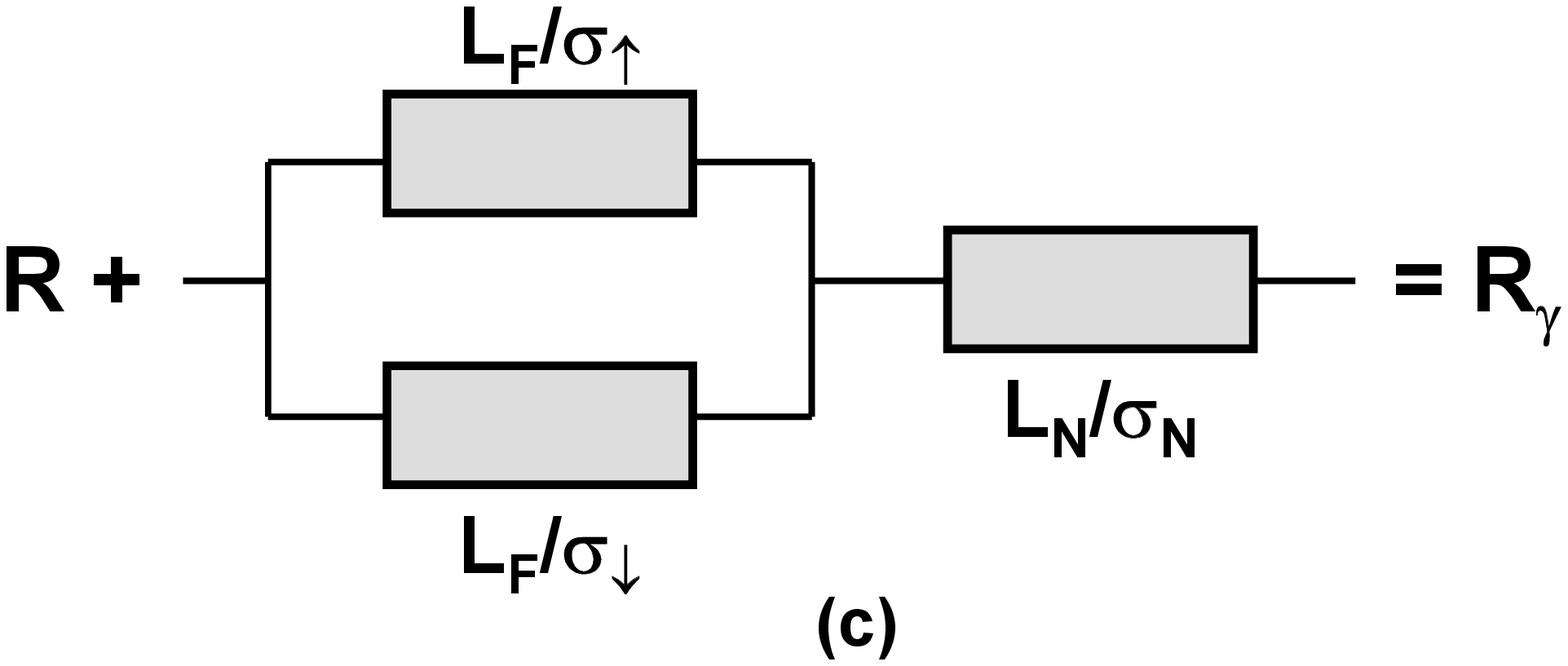}
\end{tabular}
\caption{F-N-junction ($a$) and a resistor network for it
($b$). In ($a$), F, T, and N are standing for a ferromagnet,
tunnel barrier, and normal conductor, respectively. In ($b$),
upper and lower parts of the network display effective
resistances of two parallel channels, one for up-spin and one
for down-spin electrons; see text for details. This network
provides correct expression for the spin injection coefficient
$\gamma$, Eq.~(\ref{eq2}). However, its resistance $R_\gamma$
differs from the actual resistance of the junction,
Eq.~(\ref{eq7}). Resistances $R$ and $R_\gamma$ are related by
the graphical equation shown in ($c$), see Eq.~(\ref{eq8}). }
\label{fig:resist}
\end{figure}


%

\end{document}